\UseRawInputEncoding

\documentclass{article}

\usepackage{blindtext} 

\usepackage[sc]{mathpazo} 
\usepackage[T1]{fontenc} 
\usepackage{microtype} 

\usepackage[english]{babel} 

\usepackage[hmarginratio=1:1,top=32mm,columnsep=20pt]{geometry} 
\usepackage[hang, small,labelfont=bf,up,textfont=it,up]{caption} 
\usepackage{booktabs} 

\usepackage{lettrine} 

\usepackage{enumitem} 
\setlist[itemize]{noitemsep} 

\usepackage{abstract} 

\usepackage{titlesec} 
\renewcommand\thesection{\Roman{section}} 
\renewcommand\thesubsection{\roman{subsection}} 
\titleformat{\section}[block]{\large\scshape\centering}{\thesection.}{1em}{} 
\titleformat{\subsection}[block]{\large}{\thesubsection.}{1em}{} 

\usepackage{fancyhdr} 
\usepackage{titling} 

\usepackage{hyperref} 
\usepackage{comment}
\usepackage{graphicx}

\pretitle{\begin{center}\Huge\bfseries} 
\posttitle{\end{center}} 
\title{It Remains a Cage: Ionization Tolerance of C$_{60}$ Fullerene in Planetary Nebulae} 
\author{%
\textsc{SeyedAbdolreza Sadjadi \& Quentin Andrew Parker}\\
\normalsize Laboratory for Space Research, Department of Physics,\\
\normalsize Faculty of Science, The University of Hong Kong, Hong Kong (SAR), China \\ 
\normalsize \href{mailto:ssadjadi@hku.hk \& quentinp@hku.hk }{ssadjadi@hku.hk \& quentinp@hku.hk } \\ 
\\
\normalsize \textbf{Peer-reviewed paper accepted for publication on Fullerenes Nanotubes and Carbon Nanostructures} \\ 
}
\date{\today} 
\renewcommand{\maketitlehookd}{%
\begin{abstract}
\noindent
We demonstrate that by combining two robust theoretical quantum chemistry calculation techniques, stepwise ionization of C$_{60}$ fullerene by UV and extreme UV photons can in principle occur up to a limit as high as $q=+26$ before coulomb explosion of the cage.  Furthermore, these highly ionized forms exhibit a comparable structural and bonding stability as for the neutral fullerene. Certain astrophysical sources like the central stars of planetary nebulae and the hottest white dwarf stars have sufficiently hard UV radiation fields that can result in a series of highly charged C$_{60}^{q+}$ species from $q=1$ up to $q=16$. Harsher environments, like hot X-ray bubbles in planetary nebulae, X-ray binaries and other sources, may further push the ionization right up to the $q=+26$ limit. These remarkable theoretical findings add new avenues to complex ion/molecule reactions, the chemistry of fragmentation products and additional pathways for spreading carbon throughout the universe. The implications for the emerging field of astrochemistry of C$_{60}$ fullerene in all its possible states could be profound. 

\end{abstract}
}

\begin{document}

\maketitle


\section*{Letter to Editor}

The discovery of the C$_{60}$ "fullerene" molecule, first in the laboratory by Kroto et al. \cite{Kroto1985} when attempting to mimic conditions in AGB stars and subsequently in space itself\cite{Kroto1985} where their firm identification as carriers for some of the mysterious Unidentified Infrared Emission bands (UIEs)\cite{Cami2010} and recently two Diffuse Interstellar bands (DIBs)\cite{CampbellHolzGerlichEtAl2015} was made, has led to a fundamental shift in our understanding of large carbon based molecules in space. However, many mysteries remain concerning the abundance, creation and destruction of C$_{60}$ and C$_{70}$ in the universe and the role 
they play in complex, molecular astrochemistry. 

We now demonstrate theoretically that stepwise ionization of C$_{60}$ fullerene by UV and extreme UV photons, as emitted by certain astrophysical sources like white dwarf stars and the central stars of planetary nebulae, can result in a series of highly charged C$_{60}^{q+}$ species from $q=1$  up to $q=16$. Harsher environments like hot X-ray bubbles\cite{SchonbernerSteffenWarmuth2006,Guerrero2002,LeahyZhangKwok1994} in planetary nebulae may further push the ionization right up to the $q=+26$. Furthermore, they exhibit a comparable structural and bonding stability as for the neutral fullerene. These remarkable findings potentially add new avenues to possible complex ion/molecule reactions\cite{PetrieBohme2000} and so to the emerging field of the astrochemistry of C$_{60}$ fullerene in all its possible states\cite{Woods2020}. A Positive charge of $q=+26$ is the limit that the C$_{60}$ cage can survive based on our calculations.   

Some of these cationic species up to $q=+12$ \cite{BhardwajCorkumRayner2003} have already been produced and studied with various methods in the gas phase under earth-bound laboratory conditions, e.g.\cite{PetrieBohme2000,Senn1998,BhardwajCorkumRayner2003,Pogulay2004,Sahnoun-2006-b,Kern-2013,Kern-2014}.The observed stability of the "buckyball" cage was anticipated even up to an ionization level of $q=+14$ via a robust theoretical approach.\cite{Diaz-Tendero2005,Diaz-Tendero-2005-b} 

In this work we use the latest theoretical and computational ideas for determining both the structure and form of molecular bonding within C$_{60}$ that are in agreement with findings of cited experimental and theoretical works. We find that by increasing the ionization of C$_{60}$ fullerene, that can in principle be achieved simply by winding up the UV radiation field, a range of cage like structures are produced, each composed of different CC bonding types up to and including C$_{60}^{26+}$! These gradually gain  aliphatic character as the cage is subjected to more positive charge via increasing levels of ionization. Creation and survival of such highly charged C$_{60}$ cations  can shed light on the types of molecular fragments (aliphatic/aromatic chains or rings) that get released and even upon destruction of the cage at any of the mentioned cationic forms\cite{YamazakiNakamuraNiitsuEtAl2014} or these fragments combination with other interstellar species such as H$_{2}$, CH$_{4}$, NH$_{3}$, and etc.\cite{PetrieBohme2000}

\begin{figure}
	\centering
	\includegraphics[width=1.0\linewidth, height=0.4\textheight]{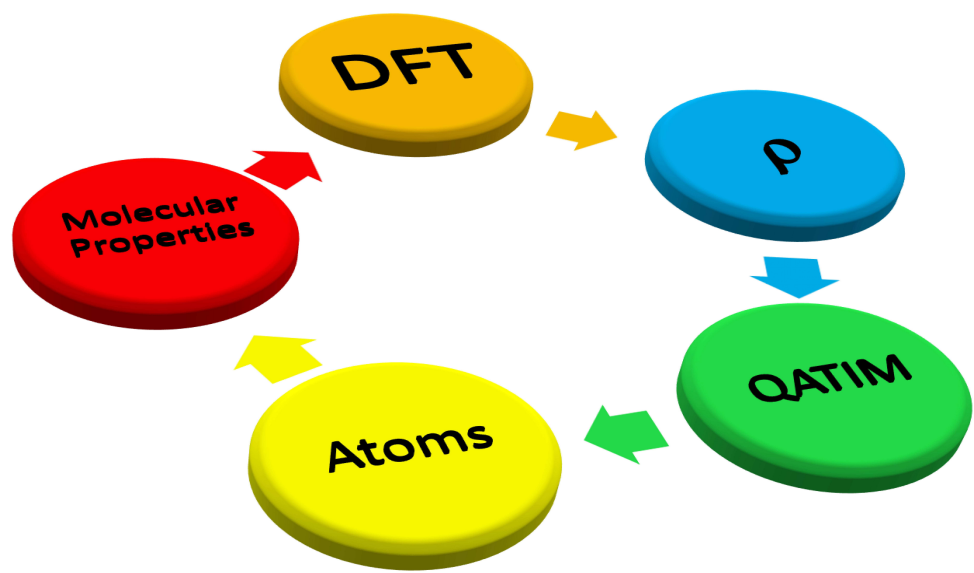}
	\caption[C60]{Relationship between DFT and QTAIM. Molecular electron density $\rho$ generated by DFT approach (Kohn-Sham method) is used to define topological atoms of chemistry and corresponding molecular properties. These data can be used to generate new functionals in DFT.\cite{Matta2017}}
	
	\label{fig1:DFT-QTAIM}
\end{figure}

Our calculations of molecular properties in this work are made possible by combining two complementary theoretical approaches: i) application of density functional theory (DFT)\cite{Kohn1999} and ii) utilization of the modern quantum theory of atoms in molecules (QTAIM).\cite{Bader1994,Matta2017} The inter-relationship between these two approaches is depicted in Figure \ref{fig1:DFT-QTAIM} 
 
In the classical approach when removing electrons from C$_{60}$ the nucleus to nucleus coulomb repulsion increases until a phenomenon known as coulomb explosion destroys the cage. The limit of charge, known as the coulomb limit, was set at relatively low ionization state of $6+$ first by Petrie et. al\cite{PetrieWangBohme1993}. By performing Hartree-Fock calculations, Cioslowski\cite{Cioslowski1996} showed that fullerene can tolerate larger positive charge, up to $10+$ before the cage undergoes coulomb explosion. Later, further calculations using a DFT approach\cite{Kohn1999}, predicted that the upper limit of ionization charge was $14+$\cite{Diaz-Tendero2005,Diaz-Tendero-2005-b}.
  
We now report that application of our novel, double theoretical approach allows extension of this previous limit up to a charge of $26+$ before the coulomb explosion occurs.  Our results predict the structural stability as determined by bond paths\cite{Bader1998,BaderMattaCortGuzm2004} of cationic species up to C$_{60}^{q+}$ (q=1-26) in astrophysical conditions, similar to neutral C$_{60}$. This is a considerably higher ionization state than what was anticipated before\cite{Diaz-Tendero2005,Diaz-Tendero-2005-b,Cioslowski1996,PetrieWangBohme1993}.
 
The detection of fullerene either alone\cite{Cami2010} or together with PAH molecules\cite{Sellgren2011} in astronomical sources, including in young planetary nebulae such as Tc~1 and IC418, is clear evidence of the persistence of its complex amorphous hydrocarbon precursor \cite{ScottDuleyPinho1997}. Its evident structural stability (in neutral or ionized forms) in harsh astrophysical environments increases the chance of even highly ionized versions of C$_{60}$ participating in a variety of new ion/molecule reactions\cite{PetrieBohme2000} and so to the formation of potentially diverse complex organic molecules and ions. Following this approach, a fundamental question arise: what are the changes a C$_{60}$ fullerene molecule is going through as the level of ionization is increased? We have attempted to answer this question as shown below.

\section*{Computational Details}

All density functional theory calculations have been conducted using the B3LYP/PC1 model, implemented in the Firefly quantum chemistry package version 8.2.0\cite{Granovsky}. This package is partially based on the GAMESS (US)\cite{SchmidtBaldridgeBoatzEtAl1993} source code. Geometry optimization of 26 cationic form of fullerene, C$_{60}^{q+}$ (q=1-26) has been conducted without symmetry constraint. The local minimum geometries characterized by normal mode frequency calculations, and their corresponding electron densities were then used for our associated calculations based on the quantum theory of atoms in molecules, performed via the AIMALL package, professional version 19.10.12.\cite{AIMALL}. The described procedure is captured by the relationship diagram in Figure \ref{fig1:DFT-QTAIM}. Symmetry point groups listed in Table \ref{T2-ionization} for local minimum geometries were determined via Chemcraft visualization interface package version 1.8.\cite{chemcraft}.Black body(BB) temperatures in Table \ref{T2-ionization} are estimated using the black body package implemented in Wolfram Mathematica software version 10.4.1.0.\cite{WolframResearch2016} 
Singlet electronic ground state via RB3LYP/PC1 and doublet via ROB3LYP/PC1 model were applied to C$_{60}^{q+}$ with even numbered and odd numbered $q$ respectively. These assumptions for electronic ground states are supported by a combined  gas phase infrared spectroscopy study and DFT calculations reported by Kern et. al\cite{Kern-2013,Kern-2014} for C$_{60}^{q+}$$(q=0, 1, 2, 3)$. 
The final symmetry point groups and zero point energy corrected ionization potential calculated for these four species are in excellent agreement with reported experimental and  theoretical calculations (See Table \ref{T2-ionization}).  	
In addition we report the vertical electronic excitation energy of singlet to triplet and doublet to quartet states for all of C$_{60}^{q+}$ in this work, lies within the range of $+2.39$ (C$_{60}^{13+}$) to $+44.83$ kcal/mol (C$_{60}$). Therefore confirming the singlet and doublet as the electronic ground states for local minimum geometries of C$_{60}^{q+}$ with even and odd numbered $q$ separately, at the applied level of calculations.

\section*{Structural Stability} 
   
The robust computational tools from the modern quantum theory of atoms in molecules provides electron density based criteria we can use to study the changes of selected molecular properties such as structure, bondings and size for C$_{60}$ during ionization. This theory is widely used to study different molecular systems including important astrochemical species such as H$_{3}^{+}$\cite{Sadjadi2017,SadjadiAbdzadehBehnejad2004} and carbon nanotubes\cite{Rashidi-RanjbarSadjadiShafieeEtAl2008}. It is worth mentioning that the possibility of formation of carbon nanotubes in space as another stable carbon allotrope, has previously captured attention in astrochemistry.\cite{Chen2019} 

Structural stability of all C$_{60}^{q+}$$(q=1-26)$ species upon ionization is determined by calculating their molecular graphs, a network of bond paths connecting carbon nuclei that form the topological structure of a molecule(see Figure \ref{fig2:topology}).\cite{Bader1998,BaderMattaCortGuzm2004}  We have found that all charged cationic forms of fullerene in this study share similar (but not equal)  topological structure and characteristic bonding features to the neutral form. No missed or ruptured bond paths between each carbon centers were observed in all cationic forms under our investigation.

\section*{Mechanism of Coulomb Explosion}

\begin{figure}
	\centering
	\includegraphics[width=1.0\linewidth, height=0.6\textheight]{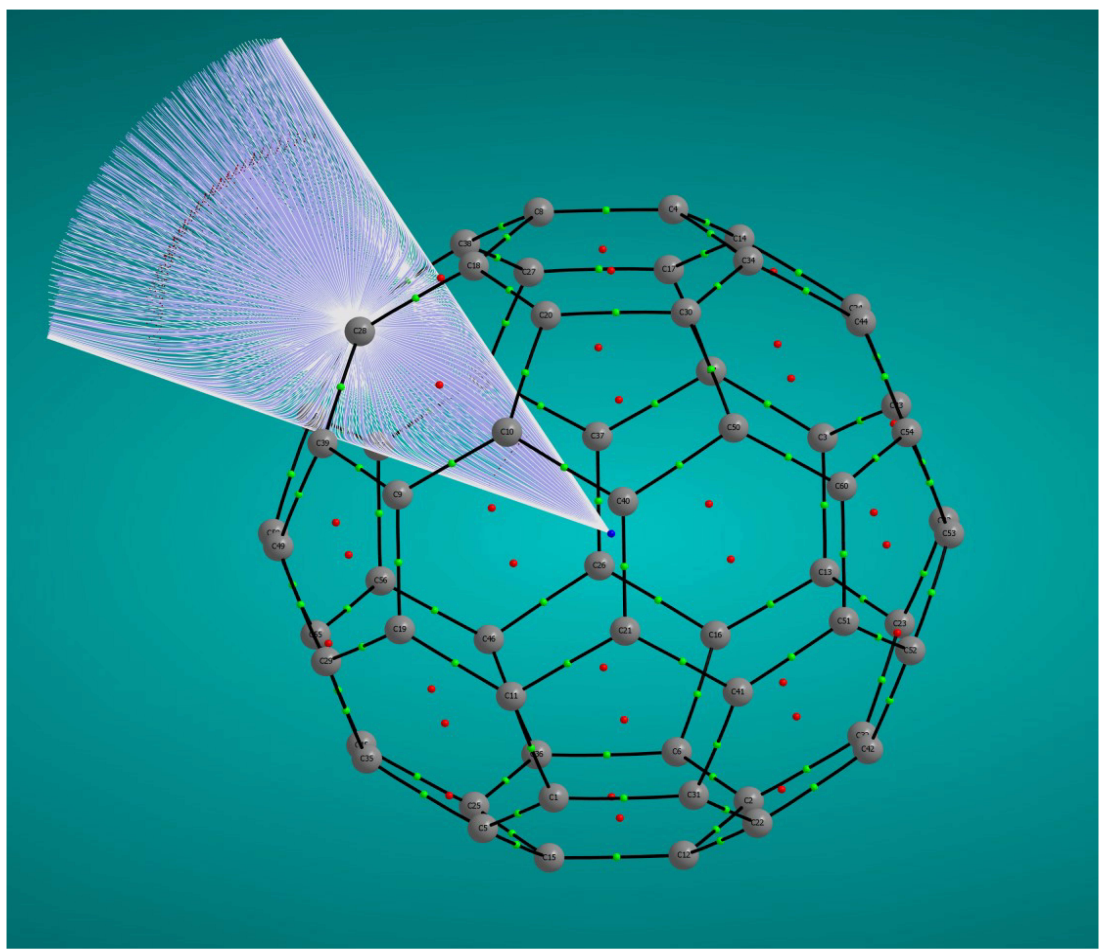}
	\caption[C60]{Topological features within the electron density $\rho$(r) of C$_{60}$ fullerene calculated by QTAIM on local minimum geometry obtained at B3LYP/PC1 level. Carbon atom nuclei (gray big circles), bond paths (black lines), bond critical points (green small circles), ring critical points (red small circle), cage critical point (blue small circle), atomic basin paths (pink lines) and interatomic surfaces (white dots).}
	\label{fig2:topology}
\end{figure}

With the well established structural stability, we then monitored the details of the structural changes in the fullerene cage via the predicted changes in the total volume of the cage, calculated as the sum of the volumes of the topological atoms (Figure \ref{fig2:topology}). These molecular properties were obtained by conducting rigorous integrations over the basins of 1560 topological atoms. These are three-dimensional slices of the molecular electron density, subject to the boundary conditions of zero flux surfaces and surround each carbon nucleus.The pyramid shape of one of these topological atoms in neutral C$_{60}$ is depicted in Figure \ref{fig2:topology}. No local minimum buckyball geometries were found by further extending the calculations from $q=26+$ to $q=30+$ indicating that no stable cage survives. \\
  
\begin{figure}
	\centering
	\includegraphics[width=1.0\linewidth, height=0.4\textheight]{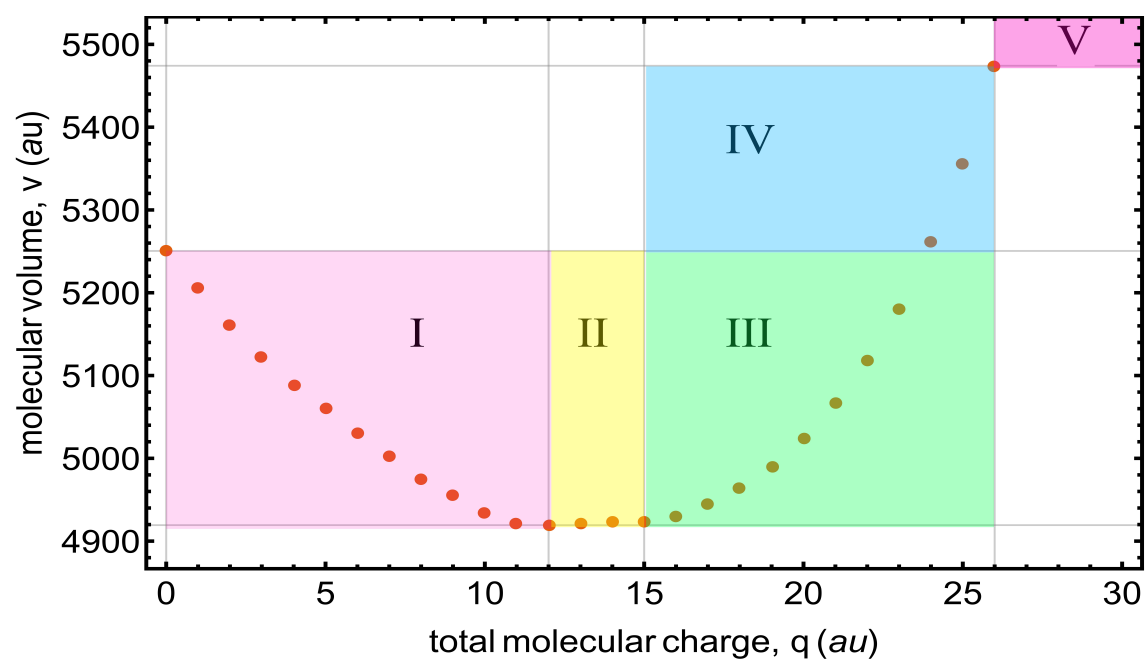}
	\caption[C60]{Molecular volume of C$_{60}$$^{q+}$ species as a function of the total charge q.   Molecular volume outer boundary is set to isodensity surface of 0.0004 (au). QTAIM calculations are conducted on B3LYP/PC1 local minimum geometries. Painted regions are labeled relative to the volume of the neutral C$_{60}$: I (contracted cage), II (relaxed cage), III (expanded cage), IV (giant cage) and V (exploded) }
	\label{fig3:volume}
\end{figure}

Figure \ref{fig3:volume} shows that according to our detailed calculations the molecular volume (v) variation as a function of charge $q$ to a good approximation a quadratic function among C$_{60}^{q+}$ species. 
Examination of our results shows we can sensibly divide this function into five distinct regions that exhibit characteristic smooth changes in molecular volume as highlighted in Figure \ref{fig3:volume} \& Table \ref{T1-volume}. The cage first contracts from the original neutral state up to a charge of $12+$, which is the most highly charged cationic fullerene currently detected in the gas phase in laboratory conditions\cite{Sahnoun-2006-b} (labeled region I or the "contracted" cage). Its size then remains effectively unchanged by further ionization up to a charge of $15+$  (region II or the "relaxed" cage). In increasing charge from $16+$ to $23+$ the cage expands relatively steeply back towards the size of the neutral form (region III, or the "expanded" cage). The cage then grows past the size of neutral fullerene and turns into a giant cage in region IV when it bears huge charges of $24+$, $25+$, and $26+$. Further ionization beyond 26 electrons finally results in cage annihilation (region V, "exploded" cage).\\
  
\section*{The Bondings}

Increasing levels of ionization has a subtle effect on both molecular structure and bonding of C$_{60}$ fullerene as it is demonstrated by changes in molecular volume. With a large number of bond paths that form each cage (Figure \ref{fig2:topology}) tracing the changes that incur to Carbon-Carbon (CC) bondings upon ionization is performed via average, minimum, and maximum delocalization indexes (DI) for each cationic species (Figure \ref{fig4:delocaliozation}). This topological electron density-dependent index carries the information on bond order and bond energy of CC bondings. For better comparison, CC bondings in benzene (C$_{6}$H$_{6}$) and ethane (C$_{2}$H$_{6}$) are chosen as a measure of the aromatic and aliphatic characters. 
Figure \ref{fig4:delocaliozation} shows that the CC bonding character as depicted by average delocalization index value, gradually shifts from average aromatic/aliphatic character for neutral C$_{60}$ (1.21 e) toward the pure aliphatic border in  C$_{60}$$^{26+}$ (1.05 e). The minimum DI values between two carbon atoms detected in C$_{60}$$^{26+}$ (0.90 e) falls slightly below the corresponding single CC value (1.0 e) in ethane. Thermal breaking of a single CC bond in ethane requires the energy of 377.4$\pm$0.8 kJ/mol\cite{CRC2004} equivalent to the ionisation that would result from a UV radiation field from a source with a BB temperature of 45000 K.     

These results demonstrate quantitatively that CC bondings are significantly strong in C$_{60}$$^{26+}$ despite the very high ionization state. This bonding strength underlies the detected structural stability of all cationic forms of fullerene.    

\section*{Implications for Astrochemistry}

These robust, theoretical, quantum chemistry results, as revealed by our novel approach, demonstrate the extraordinary flexibility of fullerene's total electron density (both pi and sigma frameworks in qualitative molecular orbital description) in relaxing to the isotopological shapes upon ionization. This can potentially extend the lifetime of fullerene and its cationic forms in space and so provides a greater chance for them to participate in other astrochemical reactions depending on the environment. Some current experimental evidence, such as the high stability of the cage structure of neutral fullerene in the solid phase under $\gamma$ ray radiation\cite{Cataldo-2009} and $\mu$s-order lifetime of cationic forms such as C$_{60}^{12+}$ in the gas phase \cite{BhardwajCorkumRayner2003,Sahnoun-2006-b} lends support to our results and suggestions.

Our quantum chemical calculations also show that the stepwise ionization of fullerene from neutral C$_{60}$ to C$_{60}^{26+}$ occurs within the wavelength range of 159.04 to 15.04 nm, i.e in the UV to extreme UV range of the electromagnetic spectrum (Table \ref{T2-ionization}). We believe that this is a viable condition within planetary nebulae with the hottest central stars and those that have been shown to possess inner X-ray bubbles\cite{SchonbernerSteffenWarmuth2006,Guerrero2002,LeahyZhangKwok1994} and also in other extreme astrophysical phenomena on stellar scales like low mass X-ray binaries \cite{ProgaKallman2002}. Thus, in the absence of any other paths to the destruction of the cage\cite{YamazakiNakamuraNiitsuEtAl2014,Biasioli1999}, the highly charged C$_{60}^{q+}$ (q=1-26) can be formed and stay within such environments.  It is well known that not only fullerene and its cationic forms but also many other highly reactive and exotic species such as C$_{2}$ can survive for a long time under astrophysical conditions.\\

\begin{figure}
	\centering
	\includegraphics[width=1.0\linewidth, height=0.4\textheight]{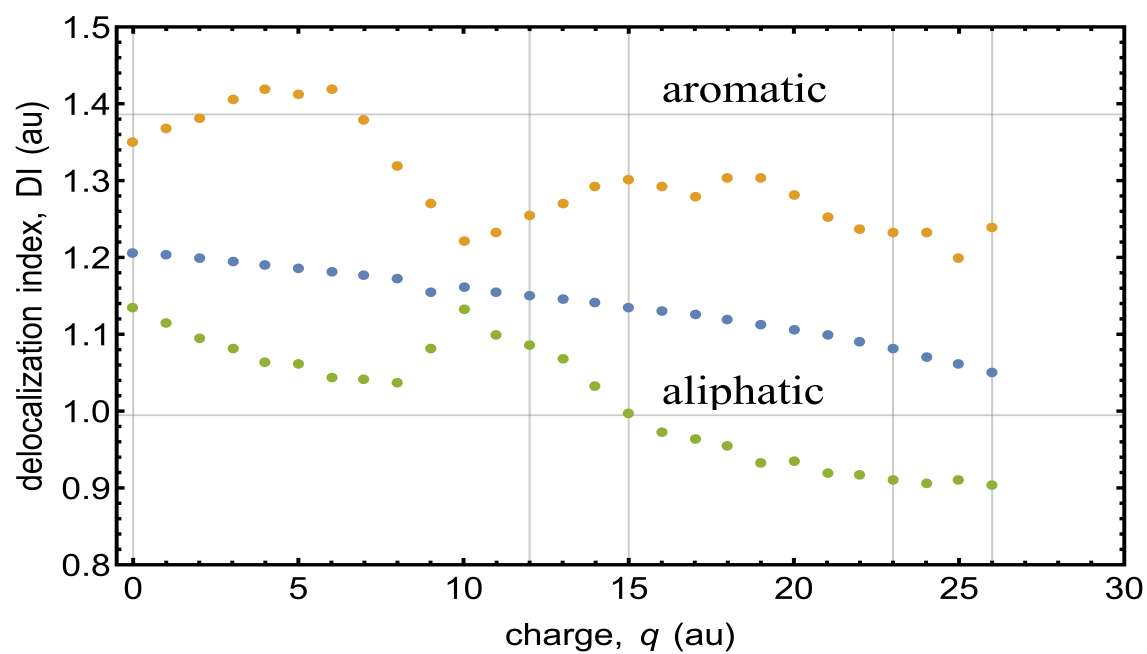}
	\caption[C60]{Carbon-carbon bonding delocalization index in C$_{60}$$^{q+}$ species as a function of the total charge q.  Aliphatic and aromatic characters are set to the delocalization indexes of CC bondings in ethane (C$_{2}$H$_{6}$) and benzene (C$_{6}$H$_{6}$) molecules respectively. Color codes: green dots are minimum DI values, orange dots are maximum DI values and blue dots are average DI values. All values from QTAIM//B3LYP/PC1 calculations.}
	\label{fig4:delocaliozation}
\end{figure}

\begin{figure}
	\centering
	\includegraphics[width=1.0\linewidth, height=0.6\textheight]{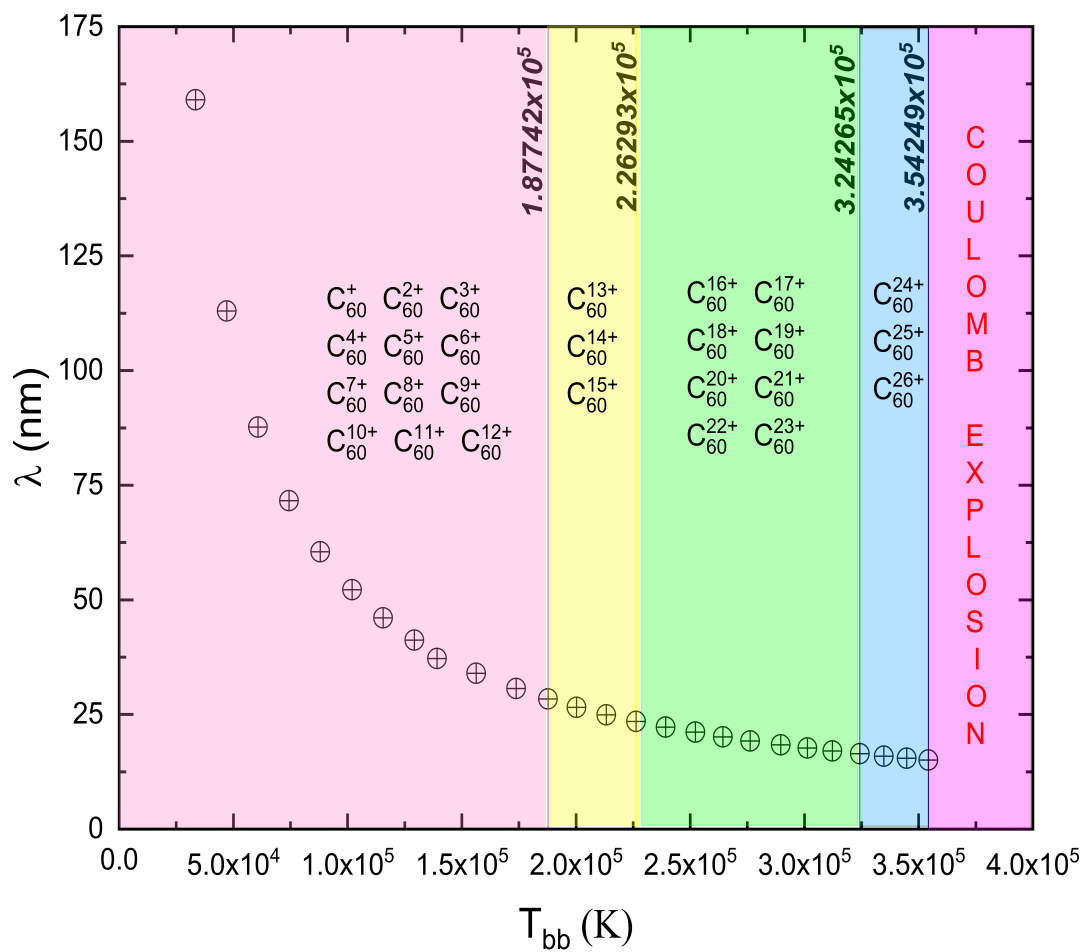}
	\caption[C60]{Theoretical ionization energy wavelength ($\lambda$) versus temperature of center star estimated by black body radiation model (T$_{bb}$) with the corresponding cationic forms of fullerene.
	Background color code of each region is similar to color code of the region depicted Figure 1. Temperature at the border of each region in bold italic. Data from Table \ref{T2-ionization}. }
	\label{fig5:centerstars}
\end{figure}

We are summarizing our findings in this study in three key points. The first is about the role of C$_{60}$ in carrying relatively large numbers of carbon atoms from the stellar furnace where it is cooked to other parts of the universe where it is consumed. Because of its demonstrated theoretical resistance to harsh temperature and radiation environments, 60 carbon atoms in some, possibly highly ionised form of fullerene, have a chance to reach a suitable environment in one bundle where they can combine with other active radicals or ions. The second point is the diversity of CC bondings as shown in Figure \ref{fig4:delocaliozation}, among  C$_{60}$ and its cationic forms. C$_{60}$ has two kind of C-C bondings, one shared between one pentagon and one hexagon (Figure\ref{fig4:delocaliozation}, minimum DI); and the other between two hexagons (Figure\ref{fig4:delocaliozation}, maxmium DI). However due to the symmetry breaking of the cage upon ionization (see Table \ref{T2-ionization}) several types of CC bonding characters can appear in between maximum and minimum DI values (Figure \ref{fig4:delocaliozation}) for different charge states.
This key chemical aspect gives each of these species their own list of typical chemical reactions such as addition and eliminations together with their specific products.
The third role that these species can play in astrochemistry is their pre-coulomb limit fragmentations (early destruction of the cage). The dominated species within the specific range of estimated temperature of center stars are presented in Figure \ref{fig5:centerstars}.
If any of these species undergo such a process, because of the CC bonding variation among them, a trail of different types of fragments are produced and spread into space. Maximum values of delocalization indexes in Figure \ref{fig4:delocaliozation} show that from neutral C$_{60}$ to C$_{60}$$^{6+}$ the fragments are more likely to contain benzene type CC bondings (associated with the aromatic character). With $q=10+$ fragments may inhere mixed aliphatic/aromatic CC bondings as shown by both maximum and minimum DI values and for $q=26+$ fragments are carrying more aliphatic CC bondings characters. Hydrogenation of each type of fragment may lead to different families of hydrocarbon molecules. \\ 
Considering these three, key astrochemical roles, C$_{60}$ can be considered as a "super spreader" of pre-biotic carbon content throughout the universe and the associated complex chemistry that may eventuate.  

\section*{Acknowledgment}
The computations were performed using research computing facilities offered
by Information Technology Services, University of Hong Kong. The authors gratefully acknowledge support from the Hong Kong Research Grants Council under grants 17326116 and 17300417.

\section*{Correspondence}
*Correspondence to: \\
SeyedAbdolreza Sadjadi (ssadjadi@hku.hk) \\
\& Quentin Andrew Parker (quentinp@hku.hk).

\clearpage

\clearpage


\clearpage

\begin{table}
	\caption{Molecular volume of the selected C$_{60}$$^{q+}$ species, calculated via QTAIM on B3LYP/PC1 local minimum geometries.}
	\centering
	\begin{tabular}{llr}
		\toprule
		\cmidrule(r){1-2}
		species & volume (au) & region$^a$ \\
		\midrule
		C$_{60}$ & 5250.463 & I \\
		C$_{60}$$^{12+}$ & 4919.335 & border I-II  \\
		C$_{60}$$^{15+}$ & 4924.137 & border II-III  \\
		C$_{60}$$^{23+}$ & 5181.204 &  III \\
		C$_{60}$$^{24+}$ & 5262.335 &  IV \\
		C$_{60}$$^{26+}$ & 5474.178 & border IV-V  \\ 
		\bottomrule
	\end{tabular}
	
	\vspace{0.1 in}
	
	{$^a$} {\footnotesize{corresponding regions depicted in Figure \ref{fig3:volume}}}
	
	\label{T1-volume}
\end{table}

\clearpage

\begin{table}
	\caption{Ionization steps lead to C$_{60}$$^{q+}$($q=1-26$) spices calculated at B3LYP/PC1 level on local minimum geometries.}
	\centering
	\begin{tabular}{llllllr}
		\toprule
		\cmidrule(r){1-3}
		 step & ESM$^a$ & CSPG$^b$ & IP (eV)$^c$ & $\lambda$ (nm) & T$_{bb}$$^d$ & spectrum region \\
		\midrule
		C$_{60}$$\rightarrow$C$_{60}^{+}$ &s$\rightarrow$d $^e$ & $I_h$$\rightarrow$$D_5d$ $^e$ &7.80$^f$ & 159.04 & 33492 & vacuum uv \\
		C$_{60}$$^{+}$$\rightarrow$C$_{60}$$^{2+}$ &d$\rightarrow$s $^e$&$D_5d$$\rightarrow$$D_5d$$^e$&10.97$^g$ & 113.00 & 47138 & vacuum uv \\
	    C$_{60}$$^{2+}$$\rightarrow$C$_{60}$$^{3+}$ &s$\rightarrow$d$^e$ &$D_5d$$\rightarrow$$C_2h$$^e$ &14.14$^h$ &  87.67 & 60753 & extreme uv \\
	    C$_{60}$$^{3+}$$\rightarrow$C$_{60}$$^{4+}$  &d$\rightarrow$s &$C_2h$$\rightarrow$$D_3d$ &17.31 & 71.63 & 74373 & extreme uv \\
	    C$_{60}$$^{4+}$$\rightarrow$C$_{60}$$^{5+}$  &s$\rightarrow$d &$D_3d$$\rightarrow$$D_3d$ &20.51 & 60.46 & 88073 & extreme uv \\
	    C$_{60}$$^{5+}$$\rightarrow$C$_{60}$$^{6+}$ &d$\rightarrow$s  &$D_3d$$\rightarrow$$D_2$ &23.74 & 52.22 & 101973 & extreme uv  \\
	    C$_{60}$$^{6+}$$\rightarrow$C$_{60}$$^{7+}$ &s$\rightarrow$d &$D_2$$\rightarrow$$C_i$ &26.92 & 46.06 & 115573 & extreme uv  \\
	    C$_{60}$$^{7+}$$\rightarrow$C$_{60}$$^{8+}$ &d$\rightarrow$s &$C_i$$\rightarrow$$D_5$ &30.09 & 41.20 & 129273 & extreme uv  \\
	    C$_{60}$$^{8+}$$\rightarrow$C$_{60}$$^{9+}$ &s$\rightarrow$d &$D_5$$\rightarrow$$D_5d$ &33.33 & 37.20 & 139173 & extreme uv \\
	    C$_{60}$$^{9+}$$\rightarrow$C$_{60}$$^{10+}$ &d$\rightarrow$s &$D_5d$$\rightarrow$$I_h$ &36.46 & 34.00 & 156273 & extreme uv \\
	    C$_{60}$$^{10+}$$\rightarrow$C$_{60}$$^{11+}$ &s$\rightarrow$d &$I_h$$\rightarrow$$C_i$ &40.43 & 30.66 & 173773 & extreme uv \\
	    C$_{60}$$^{11+}$$\rightarrow$C$_{60}$$^{12+}$&d$\rightarrow$s &$C_i$$\rightarrow$$C_i$ &43.69 & 28.38 & 187773 & extreme uv \\
	    C$_{60}$$^{12+}$$\rightarrow$C$_{60}$$^{13+}$&s$\rightarrow$d &$C_i$$\rightarrow$$C_i$ &46.68 & 26.56 & 200263 & extreme uv \\
	    C$_{60}$$^{13+}$$\rightarrow$C$_{60}$$^{14+}$ &d$\rightarrow$s &$C_i$$\rightarrow$$C_i$  &49.79 & 24.90 & 213273 & extreme uv \\
	    C$_{60}$$^{14+}$$\rightarrow$C$_{60}$$^{15+}$ &s$\rightarrow$d &$C_i$$\rightarrow$$C_i$ &52.79 & 23.49 & 226273 & extreme uv \\
	    C$_{60}$$^{15+}$$\rightarrow$C$_{60}$$^{16+}$&d$\rightarrow$s &$C_i$$\rightarrow$$C_i$ &55.79 & 22.23 & 239273$^i$ & extreme uv \\
	    C$_{60}$$^{16+}$$\rightarrow$C$_{60}$$^{17+}$&s$\rightarrow$d &$C_i$$\rightarrow$$C_i$ &58.69 & 21.13 & 252273 & extreme uv \\
	    C$_{60}$$^{17+}$$\rightarrow$C$_{60}$$^{18+}$&d$\rightarrow$s &$C_i$$\rightarrow$$C_i$ &61.61 & 20.12& 264273 & extreme uv \\
	    C$_{60}$$^{18+}$$\rightarrow$C$_{60}$$^{19+}$ &s$\rightarrow$d &$C_i$$\rightarrow$$C_i$ &64.56 & 19.20 & 276273 & extreme uv \\
	    C$_{60}$$^{19+}$$\rightarrow$C$_{60}$$^{20+}$&d$\rightarrow$s  &$C_i$$\rightarrow$$C_i$ &67.40 & 18.40 & 289673 & extreme uv \\
	    C$_{60}$$^{20+}$$\rightarrow$C$_{60}$$^{21+}$ &s$\rightarrow$d  &$C_i$$\rightarrow$$C_i$ &70.16 & 17.67 & 301273 & extreme uv \\
	    C$_{60}$$^{21+}$$\rightarrow$C$_{60}$$^{22+}$ &d$\rightarrow$s &$C_i$$\rightarrow$$C_3$ &72.85 & 17.02 & 312273 & extreme uv \\
	    C$_{60}$$^{22+}$$\rightarrow$C$_{60}$$^{23+}$ &s$\rightarrow$d &$C_3$$\rightarrow$$C_i$ &75.45 & 16.43 & 324273 & extreme uv \\
	    C$_{60}$$^{23+}$$\rightarrow$C$_{60}$$^{24+}$ &d$\rightarrow$s &$C_i$$\rightarrow$$C_i$ &77.93 & 15.91 & 334773 & extreme uv \\
	    C$_{60}$$^{24+}$$\rightarrow$C$_{60}$$^{25+}$ &s$\rightarrow$d &$C_i$$\rightarrow$$C_i$ &80.22 & 15.45 & 344773 & extreme uv \\
	    C$_{60}$$^{25+}$$\rightarrow$C$_{60}$$^{26+}$ &s$\rightarrow$d &$C_i$$\rightarrow$$C_1$ &82.44 & 15.04 & 354273 & extreme uv \\
		\bottomrule
\end{tabular}

\vspace{0.1 in}

{$^a$}{\footnotesize{Electronic Spin Multiplicity."s" stands for singlet and "d" for doublet}}.
{$^b$}{\footnotesize{Symmetry Point Group assigned to the final local minimum geometries}}.
{$^c$}{\footnotesize{Zero point energy corrected Ionization Potentials}}.
{$^d$}{\footnotesize{Black body temperature corresponds to ionization wavelength ($\lambda$)}}. 
{$^e$}{\footnotesize{In agreement with combined experimental IR spectroscopy study and theoretical calculations for electronic ground states and their corresponding symmetry point groups of C$_{60}$, C$_{60}^{+}$, C$_{60}^{2+}$ and, C$_{60}^{3+}$ in the gas phase\cite{Kern-2013,Kern-2014}}}.
{$^f$}{\footnotesize{Experimental value: 7.65$\pm$0.20 eV\cite{Pogulay2004}}}.
{$^g$}{\footnotesize{Experimental value for C$_{60}$$\rightarrow$C$_{60}^{2+}$: 18.98$\pm$0.25 eV \cite{Pogulay2004}. Our equivalent value is 18.77 eV which is the sum of the IP values of the first and second ionization steps}}.
{$^h$}{\footnotesize{Experimental value for C$_{60}$$\rightarrow$C$_{60}^{3+}$: 35.8$\pm$0.3 eV \cite{Pogulay2004}. Our equivalent value is 32.91 eV which is the sum of the IP values of the first, second and third ionization steps}}.
{$^i$}{\footnotesize{Upper limit temperature for the hottest white dwarfs}}.

\label{T2-ionization}

\end{table}

\clearpage

\clearpage

\clearpage

\clearpage

\clearpage

\clearpage

\end{document}